\documentclass[sigconf,authordraft,review=false]{acmart}

\AtBeginDocument{%
  \providecommand\BibTeX{{%
    \normalfont B\kern-0.5em{\scshape i\kern-0.25em b}\kern-0.8em\TeX}}}

\usepackage{multirow}
\usepackage{xcolor}
\usepackage{hyperref}

\def\ie{\emph{i.e.}}
\def\etc{\emph{etc.}}

\begin{document}

\title{SPEAKER VGG CCT: Cross-corpus Speech Emotion Recognition with Speaker Embedding and Vision Transformers}

\author{Alessandro Arezzo}
\affiliation{%
  \institution{University of Florence}
  \streetaddress{via Santa Marta 3}
  \city{Florence}
  \country{Italy}
  \postcode{50139}
}
\email{alessandroarezzo96@gmail.com}

\author{Stefano Berretti}
\orcid{0000-0003-1219-4386}
\affiliation{%
  \institution{University of Florence}
  \streetaddress{via Santa Marta 3}
  \city{Florence}
  \country{Italy}}
\email{stefano.berretti@unifi.it}

\renewcommand{\shortauthors}{Arezzo and Berretti}

\begin{abstract}
In recent years, Speech Emotion Recognition (SER) has been investigated mainly transforming the speech signal into spectrograms that are then classified using Convolutional Neural Networks pre-trained on generic images and fine tuned with spectrograms. 
In this paper, we start from the general idea above and develop a new learning solution for SER, which is based on Compact Convolutional Transformers (CCTs) combined with a speaker embedding. With CCTs, the learning power of Vision Transformers (ViT) is combined with a diminished need for large volume of data as made possible by the convolution. This is important in SER, where large corpora of data are usually not available.
The speaker embedding allows the network to extract an identity representation of the speaker, which is then integrated by means of a self-attention mechanism with the features that the CCT extracts from the spectrogram. 
Overall, the solution is capable of operating in real-time showing promising results in a cross-corpus scenario, where training and test datasets are kept separate. Experiments have been performed on several benchmarks in a cross-corpus setting as rarely used in the literature, with results that are comparable or superior to those obtained with state-of-the-art network architectures. 
\textcolor{black}{Our code is available at} \url{https://github.com/JabuMlDev/Speaker-VGG-CCT}
\end{abstract}




\maketitle

\section{Introduction}
The human-machine communication reached truly remarkable achievements so far. For example, 
an impressive progress has been achieved in the conversion of human speech signals in text, with a growing variety of voice assistants now available on the market for disparate uses that go from home automation to the automotive one. 
Despite of these advances, an obstacle that does not allow the man-machine interaction to be completely indistinguishable from what occurs between two people is that machines still are not ready to fully understand the emotion of the interlocutor. This constitutes a major limitation given the importance that emotions play in dialogues between two human beings. 

Motivated by these premises, recognizing the emotion of a speaker from her/his audio track is a task which is receiving an increasing attention. 
In particular, recent developments in the field of real-time SER exploited the advances in deep learning and transformed SER to an image classification task~\cite{Lech2020RealTimeSE}. In these works, the audio track is usually transformed into a spectrogram image, which is then classified using Convolutional Neural Networks (CNN) or a concatenation of CNNs and Long-Short Term Memory (LSTM) networks~\cite{Parry2019Analysis}. 
However, there are several unsolved problems that still make this task very challenging such as the lack of usable data in the desired language, the difficulty in obtaining from an audio track features that, at the same time, can be extracted in real-time and characterize the emotion, and the dependence between the representation of the emotion and the subject that expresses it. Further, most of the studies focused on the in-corpus scenario, while the more realistic cross-corpus protocol, where training and test data are completely separated has been rarely tested with. 

Developing on the above considerations, in this paper we aimed to overcome some of the aforementioned difficulties, and developed an approach capable of predicting, in real-time, the emotion of a person starting from her/his speech track. 
One main contribution of this work is that of using Vision Transformers (ViTs)~\cite{Dosovitskiy2021AnII} in place of traditional CNN for SER. ViTs proved their effectiveness in vision related tasks when used alone or in combination with CNNs, achieving performance that can surpass CNNs in many image classification tasks~\cite{Dai2021, Wu2021}, but they did not find yet application to SER. 
To this end, we used Compact Convolutional Transformers (CCTs)~\cite{Hassani2021CCT}, which represent one variant of the ViT that is simpler to train when example data are scarce. Subsequently, we extended this model to the Speaker VGG CCT that exploits the ViT self-attention mechanism in an architecture that makes use of a representation of the speakers (speaker embedding) that partially compensates for the fact that each person potentially expresses emotion in a different way. As a further contribution, we target a cross-corpus scenario. 
To validate the proposed approach, we performed a comprehensive set of experiments on the most commonly used speech datasets. First, we presented results for the challenging cross-corpus classification protocol and compared our method with state-of-the-art CNN image classification architectures fine-tuned for SER. Then, we also performed an in-corpus experiment to make our approach comparable with state-of-the-art solutions.

\section{Related Work}\label{sect:related-work}
Voice signals with associated different emotions are characterized by changes in frequency and intensity of their waveform~\cite{Beeke2009Prosody}. Therefore, several works focused on capturing those variations using different discriminating acoustic features.
Until few years ago, \textit{naive} (\ie, hand-crafted) features were proposed in various studies~\cite{ElAyadi2011pr}. These features can be categorized on the basis of various aspects such as the domain of extraction, time or frequency, timing, local or global, type, continuous or quality of the voice, \etc, using classifiers that go from Hidden Markov Models (HMMs) to Support Vector Machines (SVMs) and decision trees~\cite{Kwon2003emotion, Schuller2004, PierreYves2003}. 
Several challenges were also organized to understand the better features to use~\cite{Schuller2009interspeech, Schuller2010interspeech, Weninger2013OnTA}. 
More recently, deep learning (DL) solutions that are \textit{driven from the data} were proposed~\cite{Lieskovska2021mdpi}. Among different approaches, spectrogram-based solutions emerged as of particular interest for the fact spectrograms can be extracted quickly for real-time solutions, and can be also used as direct input to CNN and LSTM models pre-trained for the image classification task. This has the further advantage of exploiting the huge amount of available image data compared to the scarcity of speech tracks.

In the following, we focus on DL methods that used  spectrograms as input data (a summary on spectrogram generation from audio tracks is given in Appendix~\ref{sect:spectrograms}). 

\paragraph{Deep learning}
One of the first works that experimented a DL approach for SER was proposed in~\cite{Badshah2017}. In this work, spectrograms were used as features of audio data by comparing two CNN solutions: in the first one, a network was trained from scratch or with randomly initialized weights while, in the second one, a fine-tuning operation was performed on an Alexnet~\cite{Krizhevsky2012nips} pre-trained on the Imagenet dataset~\cite{Russakovsky2014, Russakovsky2015ImageNet} for the image classification task. 
In~\cite{Stolar2017icspcs}, the first work was presented that applied CNNs for real-time SER. This study compared two approaches based on the extraction of RGB spectrograms of the audio data, and on fine-tuning an AlexNet taken as a reference CNN. 
In the first architecture, named AlexNet-SVM, a pre-trained CNN was used to extract from the spectrograms, the features generated by its second fully connected layer, which are then given as input to an SVM classifier. In the second approach, instead, a fine-tuning was applied to the AlexNet (FTAlexNet) to take advantage of what the model learned to detect during pre-training, while adapting, at the same time, the weights of the last layers to the domain represented by spectrograms. The FTAlexNet shown better performance. 
This work was deepened in~\cite{Lech2020RealTimeSE}, where it is illustrated the performance decline of the models proposed, while varying the sampling frequency used to acquire the signals. 

\paragraph{SER cross corpus}
In SER, few papers focused on \textit{cross-corpus} evaluation, which is referred to as the case where datasets used for training (one or more) are different from those used in the test. 
This corresponds to a more realistic application scenario than that used by in-corpus evaluations. In these studies, increasing the data available for training was used to improve the performance. 
This was obtained either using data augmentation in the time or frequency domain, or forming a training set by the union of different datasets that can also contain examples in different languages. 

In~\cite{Parry2019Analysis}, 
a CNN, an LSTM, and an architecture given by their combination were used in a cross-corpus scenario to predict the emotions from spectrograms given as input representation of the audio tracks. 
In order to evaluate the performance, the IEMOCAP~\cite{Busso2008IEMOCAP}, EMOVO~\cite{Costantini2014EMOVO}, EmoDB~\cite{Burkhardt2005EMODB}, EPST~\cite{Liberman2002EST}, RAVDESS~\cite{Livingstone2018RAVDESS}, SAVEE~\cite{Wang2010SAVEE} and TESS~\cite{Dupuis2011Recognition} datasets were used in a three-class classification task, where each emotion label is mapped into the positive, negative or neutral class on the basis of its membership in the target domain. 
In a first experiment, the models were trained on IEMOCAP, the largest dataset among those considered, and evaluated on each of the remaining sets. In a second one, the models were in turn trained using $n$-1 datasets and tested on the remaining one. This allowed authors to evaluate the enhancement deriving from a cross-corpus training. 
The accuracy ranged from 30\% to 50\% for the first experiment, while in the second one the accuracy was around 50\% in all cases but Emo-DB and SAVEE, where it reached 69.7\% and 72.7\%, respectively. 
The best performing model was obtained by concatenating the CNN and LSTM, that resulted slightly better than the CNN model, and much better than the LSTM one.

\paragraph{Attention in SER}
One current trend in DL is that of empowering the proposed architectures with attention mechanisms. An overview of DL-based methods in SER, with specific analysis of variants that use attention mechanisms is given in~\cite{Lieskovska2021mdpi}. 

In~\cite{Moine2021speaker}, the authors proposed a self-attention mechanism to reduce the dependence of the emotion on the speaker who expresses it. To this end, a particular architecture known as ACRNN, was presented for the first time in SER in~\cite{Chen2018}. This receives in input 3D spectrograms generated by calculating the Mel spectrograms of each audio track and their respective first and second derivatives. Such 3D-representations were computed by a 3D CNN based on the application of 3D convolutions. The outputs of this network were subsequently processed by a Bidirectional LSTM (BLSTM). The output of this module, which represents a sequence with the temporal dependencies extracted from the first part of the architecture is then processed by a level of self-attention that returns the descriptor used by a fully-connected layer to classify the emotion. 
The architecture was tested with an in-corpus protocol on the Att-HACK~\cite{Moine2020AttHACKAE} and IEMOCAP datasets, and compared with the basic model given by the ACRNN. Promising results were reported with the accuracy passing, respectively, on Att-HACK and IEMOCAP, from 66.4\% and 86.4\% for ACRNN, to 88.9\% and 96.1\% for the proposed architecture.

\textcolor{black}{In~\cite{CTL-MTNet2022}, a self-attention mechanism embedded into a Capsule Network, and a Transfer Learning based Mixed Task Net (CTLMTNet) were proposed to deal with both the single corpus and cross-corpus SER tasks simultaneously. This work shares the self-attention idea with our proposal but without using ViT and speaker embedding. Moreover, in their cross-corpus tests only one dataset was used for training, while we trained our proposed network on the union of different datasets.}

\begin{figure*}[!ht]
  \centering
  \includegraphics[width=0.27\linewidth]{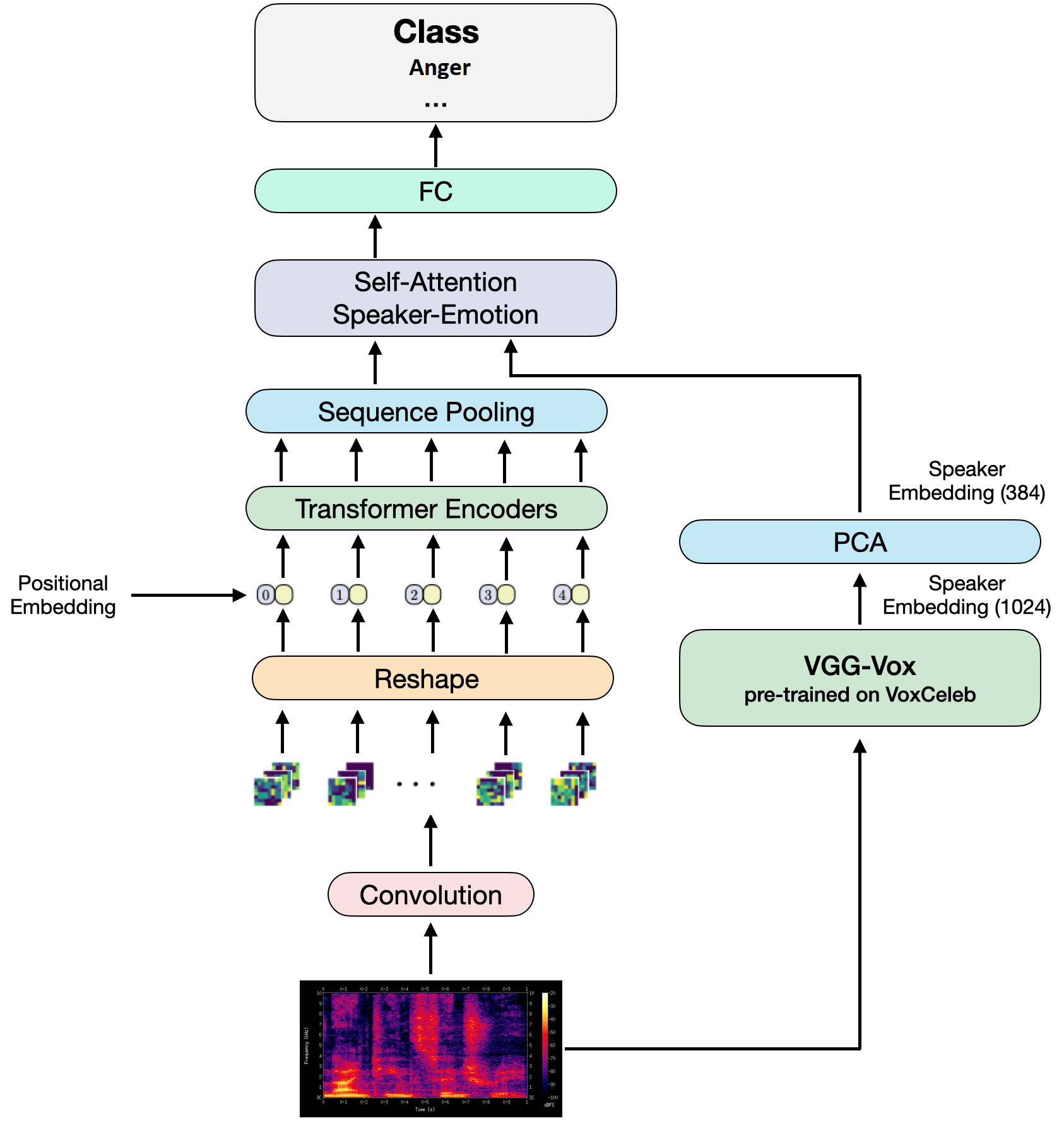} 
  \hfill
  \includegraphics[width=0.27\linewidth]{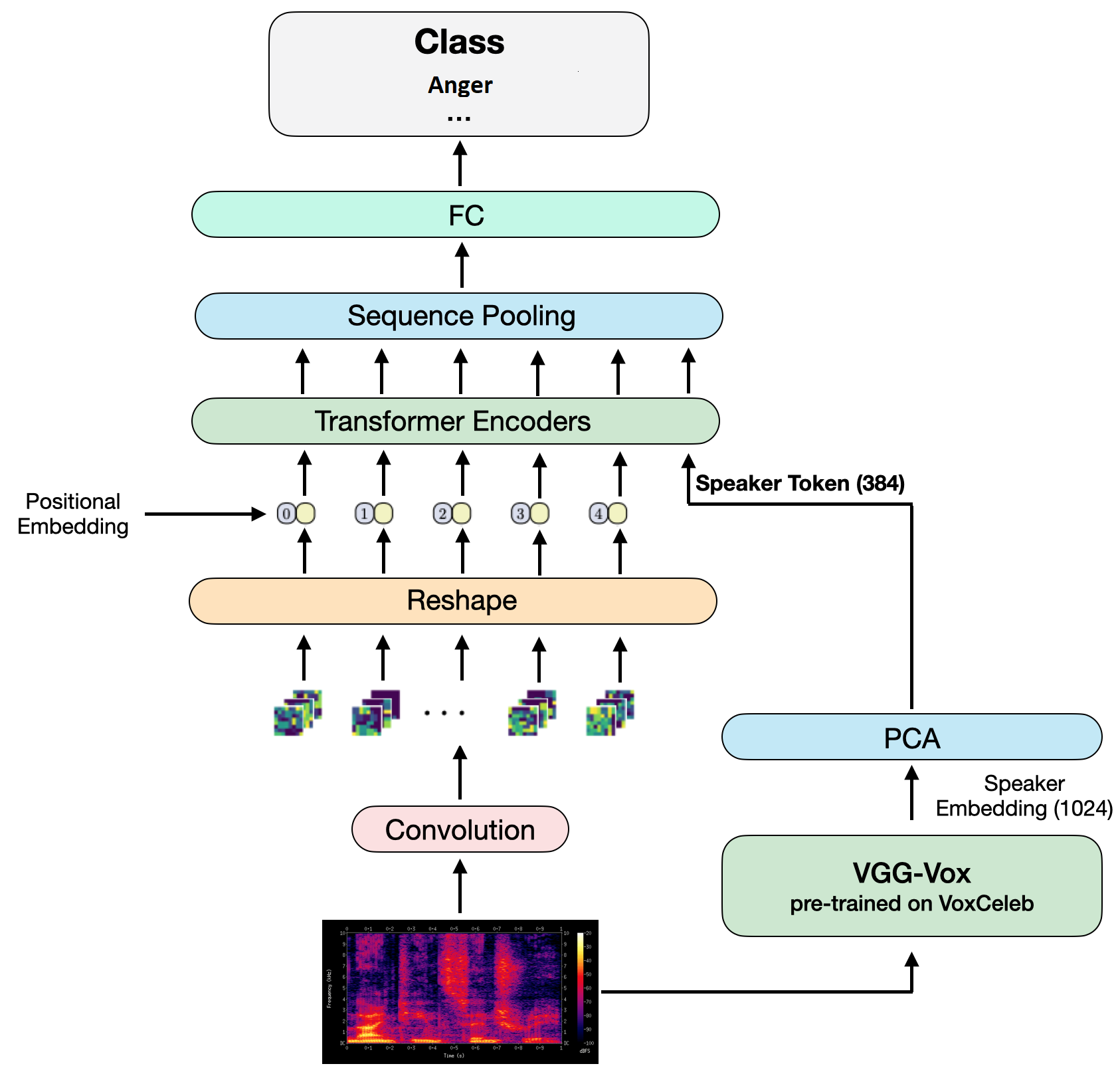}
  \hfill
  \includegraphics[width=0.27\linewidth]{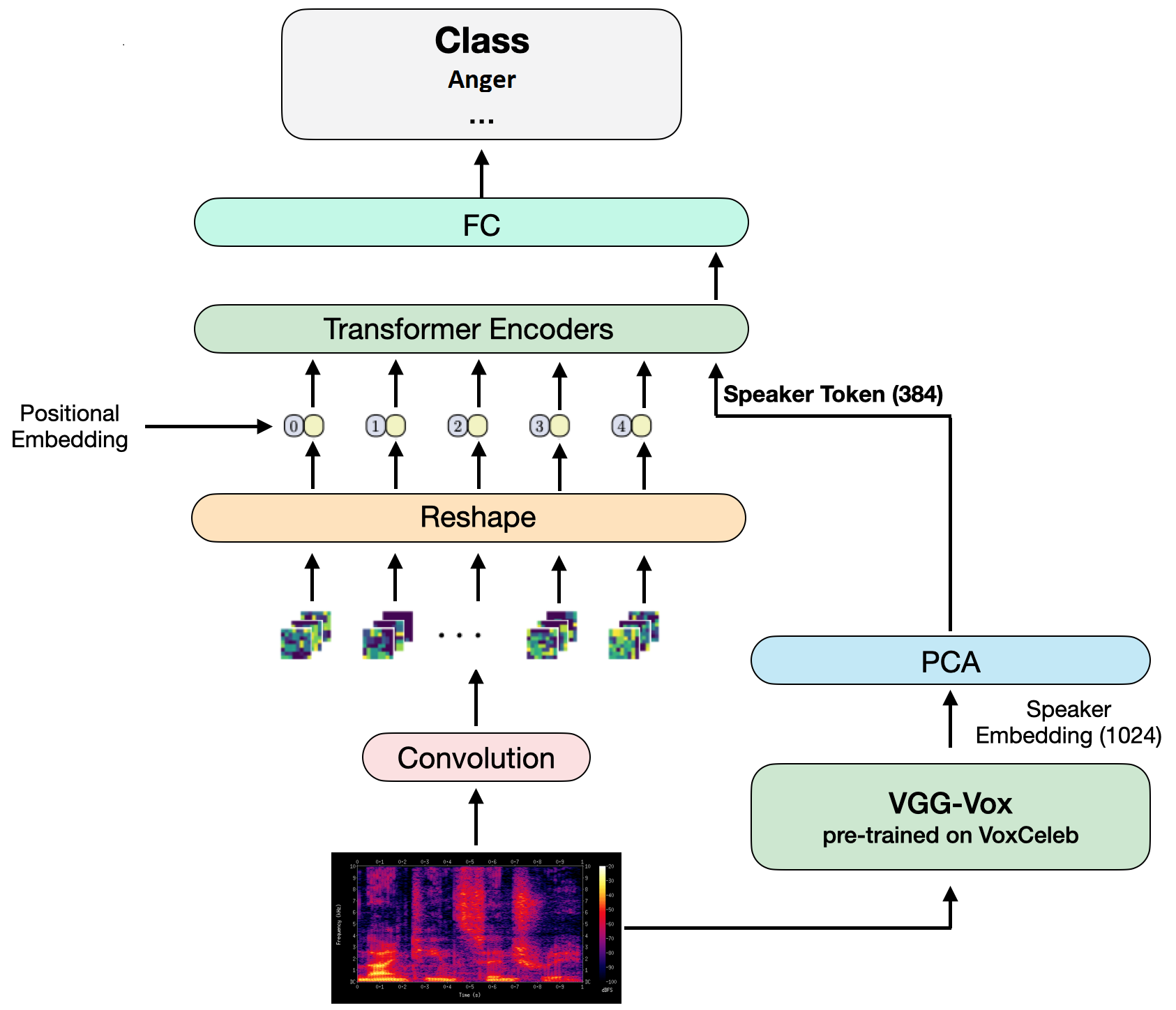}
  \caption{The proposed Speaker VGG architectures with: (left) CCT; (middle) CCT end-to-end; (right) CCT speaker token. \label{fig:speaker-vgg-cct}}
    \Description{Proposed network architecture comprising the input spectrogram image, the transformer CCT architecture with the VGG speaker embedding.}
\end{figure*}

\section{Proposed approach} \label{sect:method}
Following some recent works~\cite{Lech2020RealTime}, we propose a real-time SER method that first transforms the speech signal to a spectrogram image, then trains a neural network architecture to classify the spectrograms. 
Inspired by recent results on image analysis, we propose a Vision Transformer (ViT)~\cite{Kolesnikov2021ViT} based architecture. We did not find in the literature any previous attempt of using ViTs for SER. One reason for this is that in the image domain such architecture showed superior image classification results with respect to state-of-the-art ResNet-based models~\cite{He2016ResNet} when trained on very large amount of data. 
Since in SER there is not the same data abundance as in the image domain, we started from the Compact Convolutional Transformer (CCT)~\cite{Hassani2021CCT} that introduces convolution in ViT to make their performance superior to those of convolutional architectures even when trained on smaller amount of data. 

\subsection{Speaker VGG CCT}\label{sect:VGG}
In our solution, 
we used a CCT-14/7x2 model trained on ImageNet~\cite{Krizhevsky2012nips}, where $14$ is the number of filters in the Transformer Encoder, and $2$ is the number of convolutional layers each with filters of size $7 \times 7$. These layers are applied before the ViTs layers and receive as an input spectrogram images of size $224 \times 224$ generated from the audio track. 

In order to adapt the CCT-based architecture to SER, we investigated the possibility to better represent the dependence of a speaker emotion and the speaker him/her-self. For example, not all people express a state of happiness or frustration in the same way, causing the spectrograms corresponding to the same spoken text and emotional state to be very different when passing from one subject to another. This can lead to overfitting, where models tend to learn how to classify emotions for subjects in the training dataset, but do not generalize well in testing on spectrograms extracted from speakers that were never seen before. 
So, our idea here is to enhance the basic CCT model with a descriptor of the speakers by combining the representation through a self-attention module.
We got inspiration from the work in~\cite{Moine2021speaker}, where a 3D Attention-based Convolutional Recurrent Neural Networks (ACRNN) was trained to classify the speakers (see also Section~\ref{sect:related-work} for more details on~\cite{Moine2021speaker}). 

We propose two architectures based on 
the VGG-Vox~\cite{Nagrani2017voxceleb} speaker embedder. This is basically a CNN based on the Visual Geometry Group (VGG)-M~\cite{Chatfield2014Vgg} network trained for the speaker recognition task starting from the spectrograms of the speech signal of the Vox Celeb dataset~\cite{Nagrani2017voxceleb}. These were extracted in the same way as for all the models presented in this paper by applying a DFT with a Hamming window of size 25$ms$ and stride of 10$ms$. The resulting embedder is a vector with dimension 1024.

The first model, denoted as Speaker VGG CCT works in a similar way as in~\cite{Moine2021speaker}. In this method, the speaker embedding extracted as reported above is combined downstream of the CCT architecture, making it possible to learn the dependence between speaker and emotion. This is illustrated on the left of Figure~\ref{fig:speaker-vgg-cct}.

In the second model, instead, we defined an end-to-end solution, called Speaker VGG CCT end-to-end, where it is exploited the fact that with CCT the self-attention module is already integrated into the classifier. Therefore, the architecture ensures that the speaker embedding is given as input to the Transformers Encoder block so that it directly learns the dependence between speaker and emotion. This solution is illustrated in the middle of Figure~\ref{fig:speaker-vgg-cct}. 
In this case, the speaker embedding is passed directly to the Transformer Encoder block of CCT. 
Similar to traditional CCTs, the image is processed by a sequence of convolutional blocks that output a set of feature maps. The speaker embedding is then linked with such feature maps and the entire sequence is given as an input to the Transformer Encoder. The output is a set of vectors equal to the number of feature maps plus the one associated with the representation of the speaker. A pooling layer is then used to obtain a single vector in the dimension of the transformer starting from the exit sequence. This output is then used by an MLP network for classification. 
Therefore, the learning of the dependence between speaker and emotion underlying the speech represented by the spectrogram occurs directly in the self-attention modules internal to the Transformer Encoder. In this way, an additional vector is returned that represents the attention to be given to each patch in interpreting the speaker embedding.

\begin{table*}[!ht]
  \caption{Main characteristics of the datasets used in the experiments. Labels: \textit{Anger} (A), \textit{Boredom} (B), \textit{Disgust} (D), \textit{Excitement} (E), \textit{Fear} (F), \textit{Happiness} (H), \textit{Neutral} (N), \textit{Sadness} (S), \textit{Surprise} (Sr)}
  \label{tab:datasets}
  \small
  \begin{tabular}{lccclccccccc}
    \toprule
    Dataset & Language & \# subj. & \# samp. & labels & Anger & Disgust & Fear & Happy & Sad & Surprise & Neutral \\
    \midrule
    EmoDB~\cite{Burkhardt2005EMODB} & German & 5F / 5M & 454 & A, B, D, F, H, N, S & 127 & 46 & 69 & 71 & 62 & -- & 79 \\
    IEMOCAP~\cite{Busso2008IEMOCAP} & English & 5F / 5M & 10,039 & A, E, H, N, S, Sr & 1,103 & -- & -- & 595 & 1,084 & 107 & 1,708\\
    RAVDESS~\cite{Livingstone2018RAVDESS} & English & 12F / 12M & 1,248 & A, D, F, H, N, S, Sr & 192 & 192 & 192 & 192 & 192 & 192 & 96\\
    SAVEE~\cite{Wang2010SAVEE} & English & 4M & 480 & A, D, F, H, S, Sr, N & 60 & 60 & 60 & 60 & 60 & 60 & 120 \\
    EMOVO~\cite{Costantini2014EMOVO} & Italian & 3F / 3M & 588 & A, D, F, H, N, S, Sr & 84 & 84 & 84 & 84 & 84 & 84 & 84\\
    DEMOS~\cite{ParadaCabaleiro2020DEMoS} & Italian & 2F / 2M & 8,236 & A, D, F, H, S, Sr & 1,477 & 1,678 & 1,156 & 1,395 & 1,530 & 1,000 & -- \\ 
  \bottomrule
\end{tabular}
\end{table*}

Given the importance of this speaker embedding vector, with a different meaning than those representing the feature maps, and getting inspiration from the original implementation of ViTs, we defined a further model, called Speaker VGG CCT end-to-end Speaker Token. 
This model differs from the previous architecture in the classification methodology. In this case, there is no pooling layer, and the last level discriminates the class associated with the sample directly from the output of the Transformer Encoder associated with the speaker embedding. Given the similarity with the ViTs, in which the input corresponding to the output used to classify was defined as \textit{class token}, we denoted the speaker embedding provided by the PCA as a \textit{speaker token}. In this way, we used this particular vector that somehow represents the dependence between the speaker embedding and all the feature maps extracted from the convolutional blocks of the CCT.

\subsection{Speaker Embedding}
\begin{figure}[!ht]
    \centering
    \includegraphics[width=\linewidth]{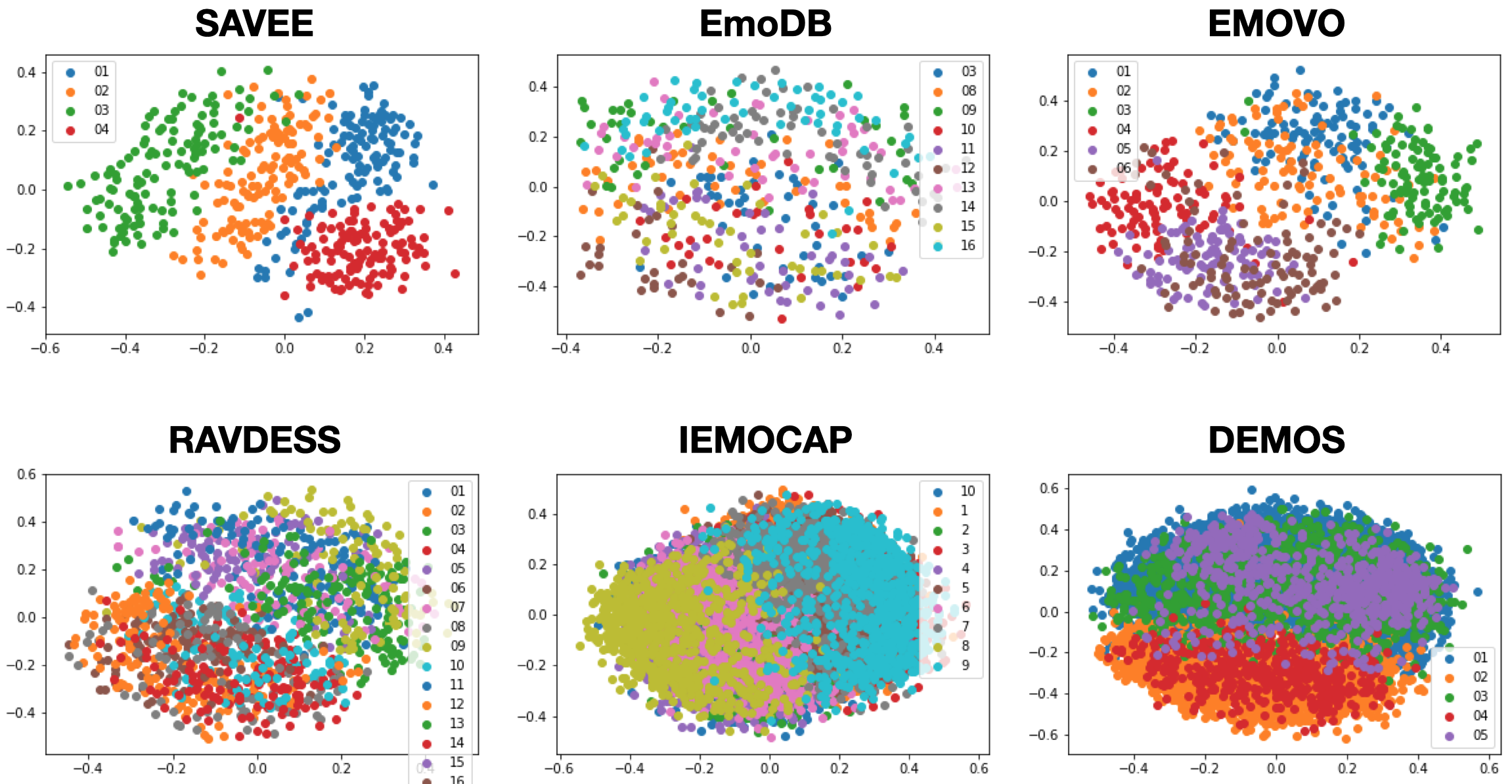}
    \caption{Plots illustrate, for each dataset, the speaker embeddings with 1,024 elements using MDS in a 2D plane, emphasizing the separation for different subjects. \label{fig:speaker-embedding}}
    \Description{.}
\end{figure}

Our architecture extracts a representation of the speakers from each voice signal with a VGG-Vox network~\cite{Nagrani2017voxceleb}, and a subsequent combination of such descriptors with information detected from spectrograms. To evidence this, for each set described in Section~\ref{sect:validation} and Table~\ref{tab:datasets}, first, the embeddings associated with each audio track were generated, thus resulting into vectors of 1,024 elements extracted from the last fully-connected layer of the VGG-Vox network. Then, Multi-dimensional Scaling (MDS) was used to represent the embeddings in a planar domain as illustrated in Figure~\ref{fig:speaker-embedding}.

We note the dimension of the speaker embedding vector was larger than the internal size of the used CCTs. In addition, the self-attention module used in each of the proposed architectures requires an equal size of the input features. So, PCA was applied to reduce the speaker embedding size to 384 elements. In this way, the desired size was obtained, while keeping the most discriminatory features among the 1,024. PCA effectively allowed us to reduce the size of the embedding, while keeping its discriminatory capability of the subjects. 

\section{Experimental Validation} \label{sect:validation}
We performed a comprehensive set of experiments to validate the proposed architectures in a real-time cross-corpus scenario. We selected the ResNet50 network pre-trained on Imagenet as the reference model to compare with. ResNet is a state-of-the-art residual architecture for image classification. The computational cost associated to its 50 layers is similar to that of the used CCT-14/7x2. 

\paragraph{Datasets}
To verify the cross-corpus performance of the proposed methods, we considered some of the most used datasets in SER as summarized in Table~\ref{tab:datasets}. All the sets taken into consideration contain examples for the seven archetypal emotions or for a subset of them, except for IEMOCAP and EMODB, which also include a few examples for the emotional states associated with \textit{boredom} and \textit{excitement}. 
Given this categorization, we restricted SER to four classes, namely, \textit{Anger} (A), \textit{Happiness} (H), \textit{Sadness} (S), and \textit{Neutrality} (N). In SER, this is often a way to reduce the difficulty of the cross-corpus task, while also reducing the problem to the most important emotions among all the primary ones. Cross-corpus tests for three classes are reported in Appendix~\ref{sect:three-classes}. 

We used the larger datasets for training, while testing was performed on the remaining sets. In particular, we considered EmoDB, EMOVO and SAVEE as test sets for cross-corpus validation, which also allowed us to evaluate the general performance for three languages, \ie, German, Italian and English, respectively. We also note all these datasets include speech tracks of emphasized sentences associated to an emotional state reproduced by actors.

\paragraph{Data Balancing}
In SER, it is quite common that audio data include many examples for certain classes and a smaller number of samples for other labels. This generally leads to the problem of \textit{class collapse}: a model tends to always predict examples with the labels associated with the classes that are more popular in the training dataset. 
This also occurs for the datasets used in this work, whose imbalance is shown in Table~\ref{tab:datasets}. It results that only the EMOVO dataset is perfectly balanced for archetypal emotions, while the others are somehow unbalanced. This problem tends to get even worse when aggregating multiple different datasets in a cross-corpus training procedure.

To address this issue and improve performance, two approaches were considered. In the first one, we implemented an undersampling technique that, for each class, removes a number of randomly selected samples equal to the difference between the samples of the aforementioned label compared to those associated with the minority class. This allows training the models on a perfectly balanced dataset; however, this significantly reduces the data available for training, thus making more severe the scarcity of data and decreasing the generalization capability of the network. 

As an alternative technique, we used a data augmentation procedure. The idea was to first apply this operation to each sample of the training set in order to obtain multiple versions associated with it. Then, for each class with fewer examples, additional samples were added in numbers equal to those missing to reach the cardinality of the majority class. 
As typical for audio computation, we first considered transformations directly applied in the time domain, specifically designed to increase a voice signal by speeding it up, slowing it down, introducing noise into it, shifting it over time, varying its tone and normalizing it. 
Another technique widely used for spectrograms derived from audio tracks consists in transforming them by acting directly in frequency. Doing so, it is possible to add vertical or horizontal bars to mask, respectively, certain moments in time rather than particular frequency bands. This allows the models to adapt less to the given training data, thus improving their generalization ability on never observed data. 
Examples of augmented spectrograms for both the approaches are reported in Figure~\ref{fig:data-aug} of the Appendix. 

\paragraph{Measures}
The following measures have been used in the evaluation, being $t_p$ and $t_n$ the \textit{true positive} and \textit{true negative}, respectively, and $p$ and $n$ \textit{positive} and \textit{negative} total:
\begin{itemize}
\item \textbf{Accuracy}: ratio between the number of correct predictions and the total number of predictions;
\item \textbf{Unweighted Average Recall} (UAR): it accounts for unbalancing of the test dataset. For the binary case, it is defined as $UAR = \frac{t_p}{p} \cdot 0.5 + \frac{t_n}{n} \cdot 0.5$, where 0.5 is replaced with $\frac{1}{classes}$ for the multi-class case. Each \textit{Recall} is weighed in the same way regardless of the number of samples of the respective class. This is considered the metric of reference in SER as the datasets are unbalanced in most of the cases;
\item \textbf{Macro-F1}: measures the quality of results for binary or multi-class classification tasks. 
It computes an average of the $F1$ measurements associated with the various classes, where each of them is defined as $F1 = \frac{t_p}{t_p + 1/2(f_p + f_n)}$.
\end{itemize}

\paragraph{Settings}
Each training was performed on a machine with an Nvidia RTX 3060 Ti GPU for a total of 50 epochs with Adam optimizer and a batch size of 32. The learning rate was selected by estimating the best performance obtained from the models on a particular test considered as a reference. 
Each architecture was trained by varying its learning rate taking as training set all datasets except EmoDB and using the latter as a test. The optimal choice of such parameter was therefore found to be $5.0e^{-4}$ for the ResNet50 and $5.0e^{-5}$ for the proposed CCT-based models.

\subsection{Cross-corpus Results}\label{sect:cross-corpus-results}
The two ``basic'' models (\ie, ResNet50 and CCT-14/7x2) were compared for the four-class task with undersampling for balancing. The IEMOCAP dataset was used for training, while tests were performed on EmoDB, SAVEE and EMOVO. We note that IEMOCAP is the largest set among those at our disposal, whereas the test datasets are those with fewer samples and characterizing three languages (\ie, English, German, Italian). Results shown in Table~\ref{tab:resnet-cct} denote a clear superiority of the CCT-based model compared to the ResNet50. 
Results also confirm the fact widely reported in the literature, that testing the models on a dataset other than the one used for training causes a noticeable performance degradation. 
\textcolor{black}{We note lower performance is obtained for SAVEE that includes less speaker compared to the others: for example, if one speaker is predicted with more difficulty this can largely affect the results.}
\textcolor{black}{We also note that IEMOCAP was used for training, though in the literature it is reported to not score good results when tested intra-corpus; we proceeded in this way because it is one of the largest dataset at our disposal and a similar choice was also proposed  in~\cite{Parry2019Analysis}.}  

\begin{table}[!ht]
  \caption{Cross-corpus results: train on IEMOCAP; test on EmoDB, EMOVO and SAVEE each with four classes (Anger, Happiness, Sadness, Neutrality). Undersampling was used to balance the IEMOCAP labels (best results in bold)}
  \label{tab:resnet-cct}
  \begin{tabular}{llccc}
    \toprule
    Dataset & Models & Accuracy & UAR & Macro-F1 \\
    \midrule
    \multirow{2}{*}{EmoDB} & Resnet50 & 44.04\% & 34.80\% & 28.76\% \\
                           & CCT-14/7x2 & \textbf{55.84}\% & \textbf{48.32}\% &  \textbf{45.53}\% \\
    \multirow{2}{*}{EMOVO} & Resnet50 & 35.12\% & 35.12\% & 27.57\% \\
                           & CCT-14/7x2 & \textbf{37.36}\% & \textbf{37.36}\% &  \textbf{30.94}\% \\
    \multirow{2}{*}{SAVEE} & Resnet50 & 27.57\% & 28.87\% & 16.86\% \\
                           & CCT-14/7x2 & \textbf{29.47}\% & \textbf{31.54}\% &  \textbf{20.62}\% \\
  \bottomrule
\end{tabular}
\end{table}

\begin{table*}[!ht]
    \caption{
    Cross-corpus results: training on IEMOCAP and DEMOS; testing on EmoDB, EMOVO and SAVEE each with four classes (Anger, Happiness, Sadness, Neutrality). 
    For each table entry, results are reported using undersampling / augmentation (us / aug) to balance the class labels (best results in bold)}
    \label{tab:resnet-speaker-cct}
    \begin{tabular}{llccc}
    \toprule
    Dataset & Models & Accuracy (us / aug) & UAR (us / aug) & Macro-F1 (us / aug) \\
    \midrule
    \multirow{5}{*}{EmoDB} & Resnet50 & 52.80\% / 48.97\% & 50.57\% / 45.20\% & 41.15\% / 39.32\% \\ 
                           & CCT-14/7x2 & 51.92\% / 54.38\% & 48.48\% / 49.00\% & 39.63\% / 38.73\% \\
                           & Speaker CCT & 53.49\% / \textbf{57.03}\% \ & 47.47\% / \textbf{52.52}\% & 37.15\% / \textbf{41.78}\% \\ 
                           & Speaker CCT end-to-end & 53.19\% / 55.16\% & 48.21\% / 50.29\% & 38.30\% / \textbf{41.78}\% \\
                           & Speaker CCT end-to-end token & 55.65\% / 54.48\% &  51.10\% / 49.30\% & 41.59\% / 39.95\% \\ 
    \midrule
    \multirow{5}{*}{EMOVO} & Resnet50 & 46.73\% / 46.53\% & 46.73\% / 46.53\% & 39.11\% / 43.78\% \\   
                           & CCT-14/7x2 & 48.91\% / 49.20\% & 48.91\% / 49.20\% & 40.89\% / 47.65\% \\   
                           & Speaker CCT & 50.29\% / 46.92\% & 50.29\% / 46.92\% & 41.64\% / 41.79\% \\  
                           & Speaker CCT end-to-end & 50.00\% / 50.79\% & 50.00\% / 50.79\% & 43.32\% / \textbf{48.79}\% \\   
                           & Speaker CCT end-to-end token & \textbf{51.69}\% / 51.39\% & \textbf{51.69}\% / 51.39\% & 43.43\% / 47.36\% \\ 
    \midrule
    \multirow{5}{*}{SAVEE} & Resnet50 & 32.67\% / 31.78\% & 37.99\% / 35.62\% & 32.39\% / 31.10\% \\   
                           & CCT-14/7x2 & 35.33\% / 33.22\% & 41.67\% / 36.88\% & 34.40\% / 32.32\% \\   
                           & Speaker CCT & 32.78\% / 33.00\% & 39.02\% / 37.85\% & 31.62\% / 31.61\% \\   
                           & Speaker CCT end-to-end & 36.66\% / \textbf{39.89}\% & 42.71\% / \textbf{45.35}\% & 35.91\% / \textbf{40.03}\% \\  
                           & Speaker CCT end-to-end token & 32.11\% / 34.33\% & 38.68\% / 39.86\% & 30.26\% / 32.87\% \\  
    \bottomrule
\end{tabular}
\end{table*}

Table~\ref{tab:resnet-speaker-cct} reports the comparison between all the proposed models. We still performed a cross-corpus experiment, with four classes and with undersampling for balancing (left values for each table entry). In this case, we used a cross validation protocol, where for each test dataset the models were trained several times, each time excluding the test set and, in round, selecting as validation RAVDESS and the remaining two datasets of test among EmoDB, EMOVO and SAVEE. 
IEMOCAP and DEMOS that include more samples among the available datasets were used for training. 
From the table, it can be noted that all the metrics are generally superior to those obtained by training on IEMOCAP only for ResNet50 and for the CCT model. This highlights the fact, already reported in the literature, that cross-corpus training can bring significant benefits in SER. As for the comparison between the proposed architectures, it is evident that the learning of the dependence between speakers and emotion allows us to improve the performance of the models also in a cross-corpus test. In particular, the end-to-end architectures get the best metrics. 
Among the proposed models, for EmoDB the one without speaker token (which uses instead the pooling layer to classify) achieves the best results, while for the other two sets it was the other architecture to achieve the best results. On average, by focusing on the two most relevant metrics, the speaker token model has obtained a higher UAR, but with a lower value for Macro-F1.

The gap with the state-of-the-art is of about 2\%. This gap increases by performing the same experiments but with training carried out on balanced datasets with data augmentation, as shown by the values reported on the right of each entry of the table. The best models are those end-to-end, with a gap between the best performing one (Speaker CCT end-to-end) compared to ResNet50 of around 6\% and 5\%, respectively, for accuracy and Macro-F1. Results using augmentation for data balancing are higher for almost every model, demonstrating that using more data is beneficial for improving the methods capability of generalizing at inference time.

With reference to the results illustrated so far, it should be noted the relevance of the used metrics. In fact, it has pointed out the two reference measures are the UAR and the Macro-F1 that can be considered as complementary: the UAR indicates the accuracy obtained on average taking into account the unbalancing of the test datasets, while the Macro-F1 measures the quality of the results. Even for a lower UAR value, if the respective Macro-F1 value is higher, the quality of the results, explained as the concentration of the values along the diagonal of the confusion matrix will be better. 
See Figure~\ref{fig:confusion-matrix} in the appendix for a confusion-matrix example using the two end-to-end models.  
\textcolor{black}{Overall, we designated the Speaker CCT end-to-end as our best model because it outperforms the others in terms of Macro-F1, and almost in all the cases for accuracy and UAR. We also point out that the accuracy values on EMODB and SAVEE do not say much because these sets are unbalanced. More relevant are UAR and Marco-F1, this latter indicating the distribution of the predictions between the different classes.}



\subsection{Intra-corpus Results}
An intra-corpus experiment was performed on the EmoDB dataset, thus making the approach comparable with other solutions in the literature. We used a Leave-One-Speaker-Out (LOSO) protocol, where iteratively a speaker is used for test and models are trained only on the samples of the rest of the subjects. 

\begin{table}[!ht]
    \caption{Intra-corpus results: LOSO protocol on EmoDB using anger, happiness, sadness, neutrality, fear, and surprise. 
    }
    \label{tab:intra-corpus}
    \begin{tabular}{lccc}
    \toprule
    Models & Accuracy & UAR & Macro-F1 \\
    \midrule
    Resnet50 & 67.21\% & 63.64\% & 60.57\% \\
    CCT & 74.58\% & 70.75\% & 68.16\% \\
    Speaker CCT & 75.00\% & 70.56\% & 67.48\% \\
    Speaker CCT end-to-end & 70.18\% & 67.04\% & 63.86\% \\
    Speaker CCT end-to-end token & 68.44\% & 66.85\% & 63.19\% \\
    \midrule
    AlexNet~\cite{Lech2020RealTimeSE}* & 80.5\% & – & – \\
    \bottomrule
\end{tabular}
*Results taken from the original paper using 5-fold cross validation
\end{table}

Table~\ref{tab:intra-corpus} shows the results for each of the proposed architectures obtained by averaging the values detected for all the speakers. The performance shows a different trend with respect to the cross-corpus experiment, with the models using the speaker embeddings showing lower accuracy and Macro-F1. This can be explained by the fact that, to work properly, such architectures need more speakers than the 10 included in EmoDB. 
In fact, such methods during the training learn the dependence between the representation of the speakers and their emotions. During inference, the models should be able to match the embedding of the tested speaker with the closest one among those in the training set. Therefore, if the speakers observed during the training are too few, such association can be inaccurate.
Notably, CCT and all the proposed variants achieved superior performance compared to RenNet50. This is inasmuch relevant as we propose for the first time the use of ViT in SER. 

The table also reports results for the method in~\cite{Lech2020RealTimeSE}, which is a reference one for the application of CNN in real-time SER. In that study, authors classified the spectrograms using AlexNet pre-trained on Imagenet. Despite the reported results were of about $80\%$, we note they used a 5-fold cross-validation protocol, where the trained models were evaluated on audio produced by speakers already viewed by the network during the training, with a significant loss of generality compared to an evaluation performed with the LOSO protocol. 

\section{Conclusions} \label{sect:conclusions}
In this paper, we addressed the SER problem in real-time and considering a concrete application scenario where most of the evaluation is performed cross-corpus. 
The proposed approach has some innovative aspects. First of all, it is the first to apply a ViT model in SER through the CCT variant, also integrating the learning of dependence between speaker embedding and emotion by means of self-attention on its own or in an end-to-end architecture.
Furthermore, looking for solution as general as possible, all proposed methods were tested in a cross-corpus setting as it is rarely the case in SER. 
Results show evident improvements with respect to ResNet50 taken as state-of-the-art architecture for real-time SER. In particular, the proposed end-to-end architecture turned out to be the best performing one. 

\begin{acks}
This work was done during a research internship of Alessandro Arezzo at QuestIT, Siena, Italy. The authors would like to thank QuestIT for the support.
\end{acks}


\appendix

\section{Spectrograms}\label{sect:spectrograms}
\begin{figure}[!ht]
    \centering
    \includegraphics[width=\linewidth]{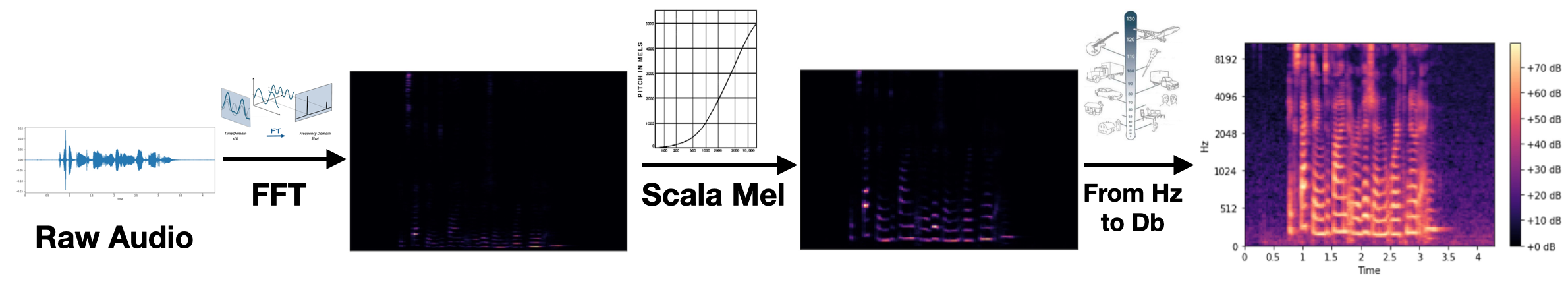}
    \caption{Generation of the spectrogram of an audio signal. \label{fig:mel-spectrograms}}
    \Description{.}
\end{figure}

The spectrogram of an audio signal is generally defined as its own representation that makes explicit the contributions of each frequency for each time frame. The process of extracting the spectrogram from an audio track is summarized in Figure~\ref{fig:mel-spectrograms}. 
First, the Fast Fourier Transform (FFT) is applied to the audio signal sampled into the time domain using small time windows; in this way, for each of them the contribution of each of its components in the frequency domain is obtained. Putting together the results obtained on all the windows, a 2D representation of the original signal is derived, where each element on the horizontal axis represents a window and the values on the vertical axis explain the amplitudes of the spectrum.
To compensate for non linearity, the Mel spectrograms are used that map each frequency into a corresponding one using the Mel scale. This, to bring the frequencies to a logarithmic scale, transforms low frequencies into a larger range than that used for the high ones, so that the coefficients of the lower frequencies are represented by a larger region on the spectrogram. 
A logarithmic scale is also used for the amplitude representation that transforms the frequencies from Hertz (hz) to Decibels (db). A final image with three well distributed channels representing the original audio signal is obtained. 

In our work, to generate the spectrogram images, a Discrete domain Fourier Transform (DFT) was first applied to the audio tracks in the time domain to obtain a frequency representation, which is then mapped to new values using the \textit{Mel scale} and \textit{decibels}. Parameters that characterize such process are the factors used for applying the DFT. Following other works in SER, we applied the aforementioned transform with a \textit{Hamming window} of $25ms$ and a stride of $10ms$. 

These descriptors are characterized by the bands frequency, represented by their number and the minimum and maximum frequencies, and aspects related to temporal sections, such as the length of each window and the stride that identifies the number of samples of which to shift the frames at each step. Usually, in SER, 128 frequency bands are selected, with a window size of 25ms and a stride of 10ms.

\section{Additional Results}\label{sect:add-results}

\subsection{Data Augmentation}\label{sect:data-augmentation}
\begin{figure*}[!ht]
    \centering
    \includegraphics[width=0.11\linewidth]{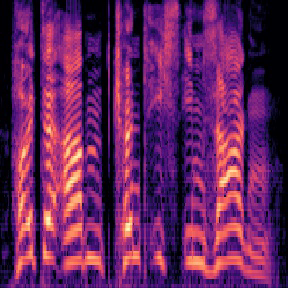}
    \hfill
    \includegraphics[width=0.11\linewidth]{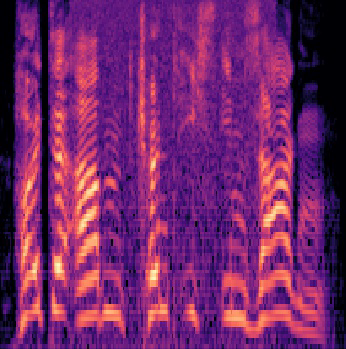}
    \includegraphics[width=0.11\linewidth]{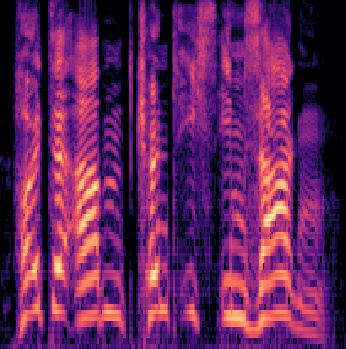}
    \includegraphics[width=0.11\linewidth]{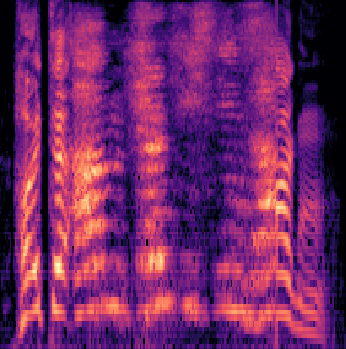}
    \includegraphics[width=0.11\linewidth]{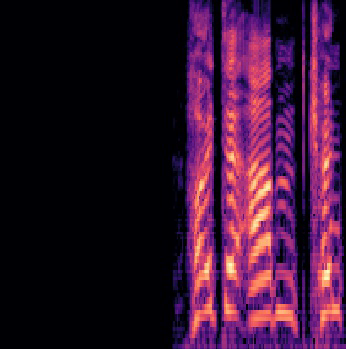}
    \includegraphics[width=0.11\linewidth]{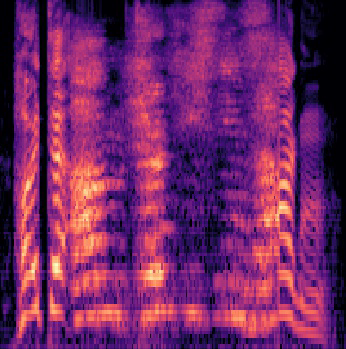}
    \hfill
    \includegraphics[width=0.11\linewidth]{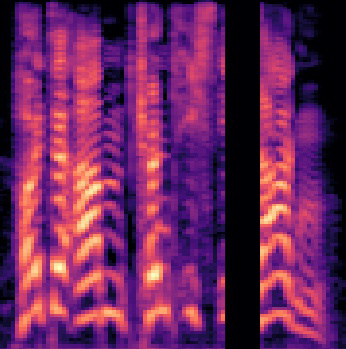}
    \includegraphics[width=0.11\linewidth]{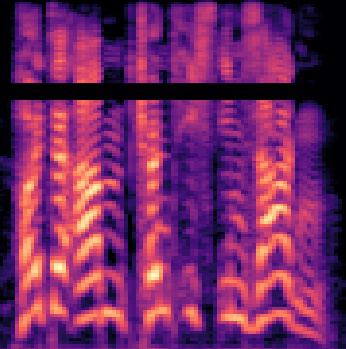}\\
    \begin{minipage}{0.11\linewidth}
    \centering (a)
    \end{minipage}
    \hfill
    \begin{minipage}{0.55\linewidth}
    \centering (b)
    \end{minipage}
    \hfill
    \begin{minipage}{0.22\linewidth}
    \centering (c)
    \end{minipage}
    \caption{Examples of spectrograms associated with data augmentation transformations applied to a sample of EmoDB. (a) Spectrogram of the original signal; (b) Five time domain transformations: noise, normalization, tone variation, shifting, and speed change; (c) Two frequency domain transformations: masks introduced in time and frequency, respectively.
    \label{fig:data-aug}}
    \Description{.}
\end{figure*}

Figure~\ref{fig:data-aug} shows, for the sample spectrograms in (a), five time domain transformations, \ie, noise, normalization, tone variation, shifting, and speed change in (b), and two transformations in the frequency domain, \ie, masks introduced in time and frequency in (c).

\subsection{Confusion Matrices}\label{sect:confusion-matrix}
Here, we further investigate the relevance of the used metrics. In Section~\ref{sect:validation}, we pointed out the two reference measures for evaluation are the UAR and the Macro-F1 that can be considered as complementary: the UAR indicates the accuracy obtained on average taking into account the unbalancing of the test datasets, while the Macro-F1 measures the quality of the results. Even for a lower UAR value, if the respective Macro-F1 value is higher, the quality of the results, explained as the concentration of the values of the confusion matrix along the main diagonal, will be better. 

\begin{figure}[!ht]
    \centering
    \includegraphics[width=0.48\linewidth]{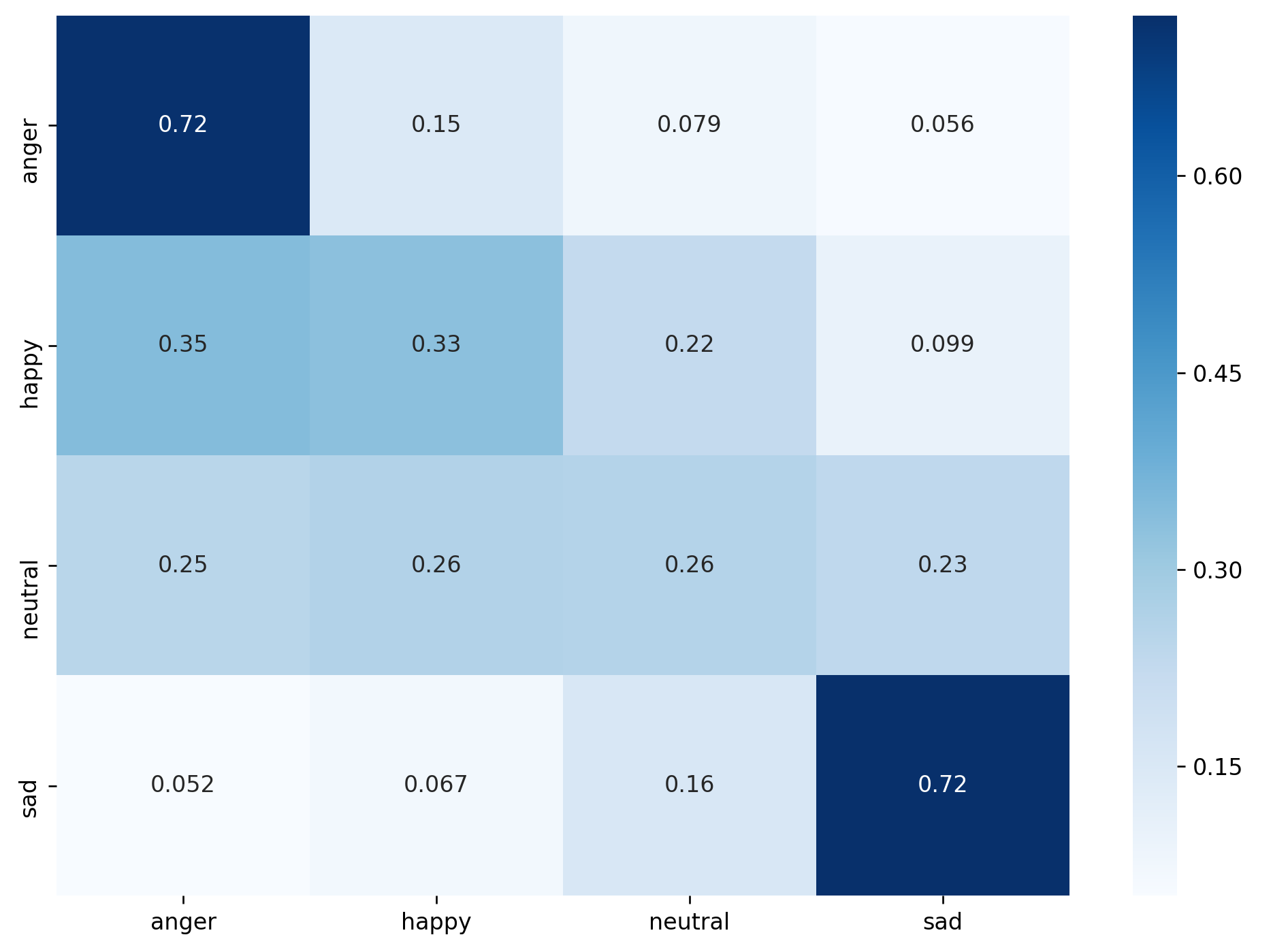}
    \hfill
    \includegraphics[width=0.48\linewidth]{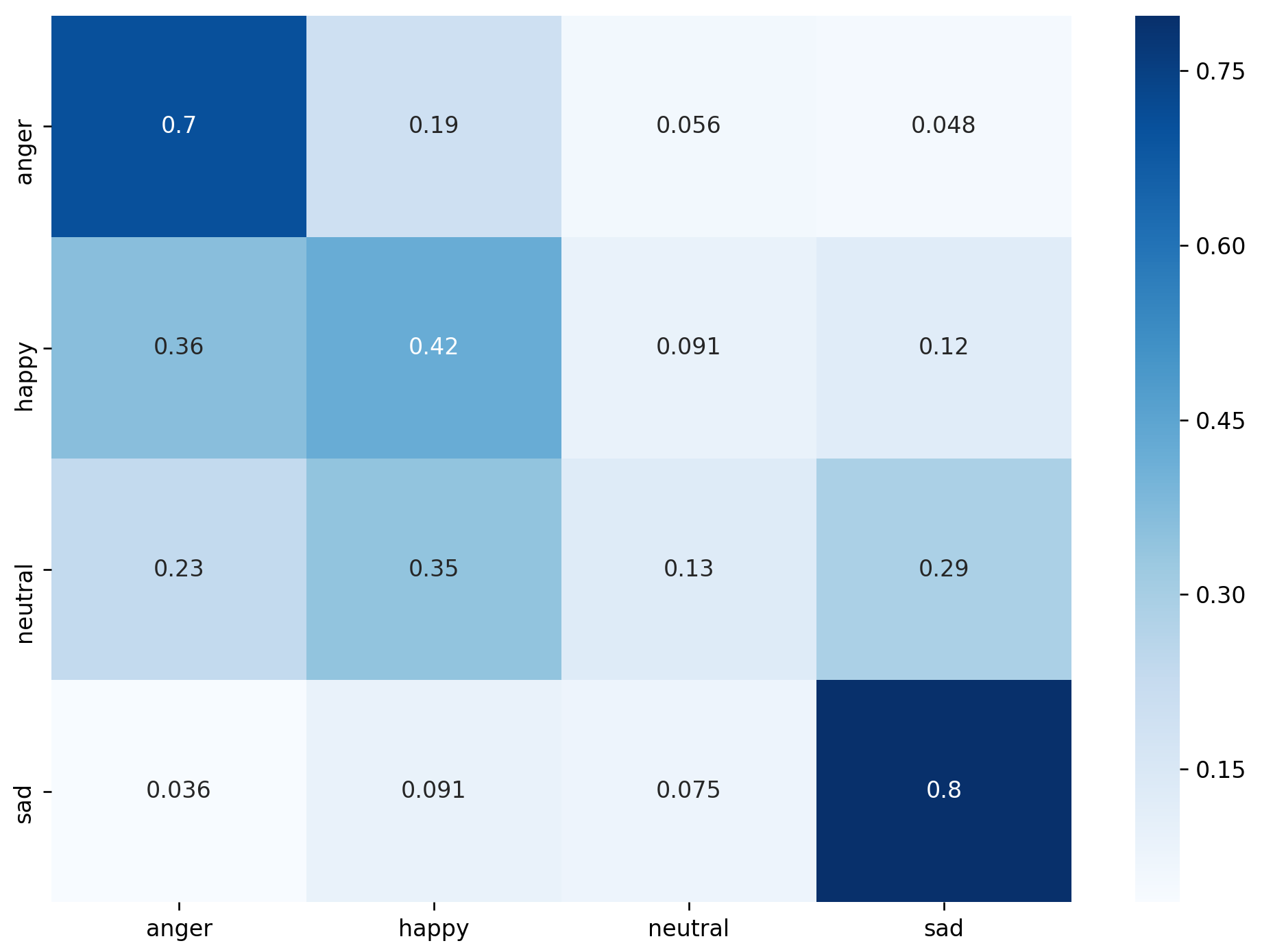}
    \caption{Confusion matrices referring to cross-corpus experiments with four classes and data augmentation for balancing. Tests are performed on EMOVO with the Speaker VGG CCT end-to-end (left), and with the Speaker VGG CCT end-to-end with speaker token (right). \label{fig:confusion-matrix}}
    \Description{.}
\end{figure}

An example of this is shown in Figure~\ref{fig:confusion-matrix}, where the matrices obtained from the two end-to-end models are shown by testing on EMOVO with the four-class cross corpus training and the balancing procedure for data augmentation. The matrix on the left and on the right refer, respectively, to the model that uses the pooling layer for the extraction of the classification features (Speaker VGG CCT end-to-end), and to the model with speaker token (Speaker VGG CCT end-to-end token).  

Despite from the results shown in Table~\ref{tab:resnet-speaker-cct}, the best accuracy is obtained with the method that uses the speaker token, the confusion matrices show a more balanced performance between classes in the Speaker VGG CCT end-to-end case. The highest quality, quantitatively expressed by a higher Macro-F1 value is highlighted by a better ability of the model to predict the class neutral, which is poorly evaluated by the other architecture. 
It is therefore evident how much a greater value of the Macro-F1 is even more indicative of that of the UAR, that is instead most often used in SER. 
In fact, for application purposes, it would be more useful to have a model that performs well enough for all classes, rather than one capable of perfectly predicting only a subset of them, while providing poor performance on the remaining ones.

\subsection{Cross-corpus Experiments on Three Classes}\label{sect:three-classes}
For three-class classification, each label of the reference datasets is mapped into \textit{positive}, \textit{negative} and \textit{neutral} emotions, according to the following association: negative--Anger (A), Boredom (B), Disgust (D), Fearful (F), Sadness (S); positive--Excitement (E), Happiness (H), Surprise (Sr); neutral--Neutral (N).
Using three classes further allows reducing the number of classes to discriminate, while increasing, at the same time, the number of examples since none of them is discarded but just remapped to another space. 
However, since there are many labels associated with the negative class, and a lower number with the positive, and only one with the neutral class, there is a notable increase in the imbalance already present in the used datasets. All this considered, the difficulty in discriminating these features still keeps high despite the lower number of classes. 

\begin{table*}[!ht]
    \caption{
    Cross-corpus experiments on the Positive, Negative, and Neutral classes, and test on the EMOVO dataset. For each table entry, results are reported using undersampling / augmentation (us / aug) to balance the class labels. (best results in bold)}
    \label{tab:three-classes}
    \begin{tabular}{lccc}
    \toprule
    \textbf{Models} & \textbf{Accuracy} (us / aug) & \textbf{UAR} (us / aug) & \textbf{Macro-F1} (us / aug) \\
    \midrule
    Resnet50 & 52.15\% / 54.24\% & 38.49\% / 34.81\% & 36.58\% / 31.34\% \\ 
    \midrule
    CCT-14/7x2 & 51.79\% / 54.48\% & 38.53\% / 35.91\% & 37.21\% / 30.79\% \\
    Speaker CCT & 50.73\% / 54.81\% & 38.96\% / 36.29\% & 37.39\% / 31.62\% \\ 
    Speaker CCT end-to-end & 53.80\% / 54.76\% & 39.14\% / \textbf{36.57}\% & 37.09\% / \textbf{32.63}\% \\ 
    Speaker CCT end-to-end token & 55.95\% / \textbf{57.14}\% & \textbf{40.04}\% / 35.09\% & 39.64\% / 29.20\% \\ 
    \midrule
    CNN~\cite{Parry2019Analysis}* & -- / -- & 39.93\% / --~~~ & -- / -- \\
    \bottomrule
\end{tabular}
\end{table*}

As evaluation pipeline, we used the same procedure as described in Section~\ref{sect:cross-corpus-results} for the cross-corpus training with four-classes. However, the evaluation was constrained to the EMOVO dataset only. Results reported in Table~\ref{tab:three-classes} show the same trend of the tests performed for four-classes, with the end-to-end models tending to have superior performance than the others.

However, unlike what was observed with the four-classes task, in this case the balancing technique with data augmentation achieves lower performance than that with undersampling. This is due to the fact that by running data augmentation, the resulting models tend to predict almost always with the negative class for any test example. 
The cause of this problem is that in this case the imbalance of the training dataset already present for the four-class task increases considerably. Indeed, according to the mapping, many labels of the reference datasets are mapped into the negative class, few in the positive one and only one in the neutral one. 

\begin{figure}[!ht]
    \centering
    \includegraphics[width=0.48\linewidth]{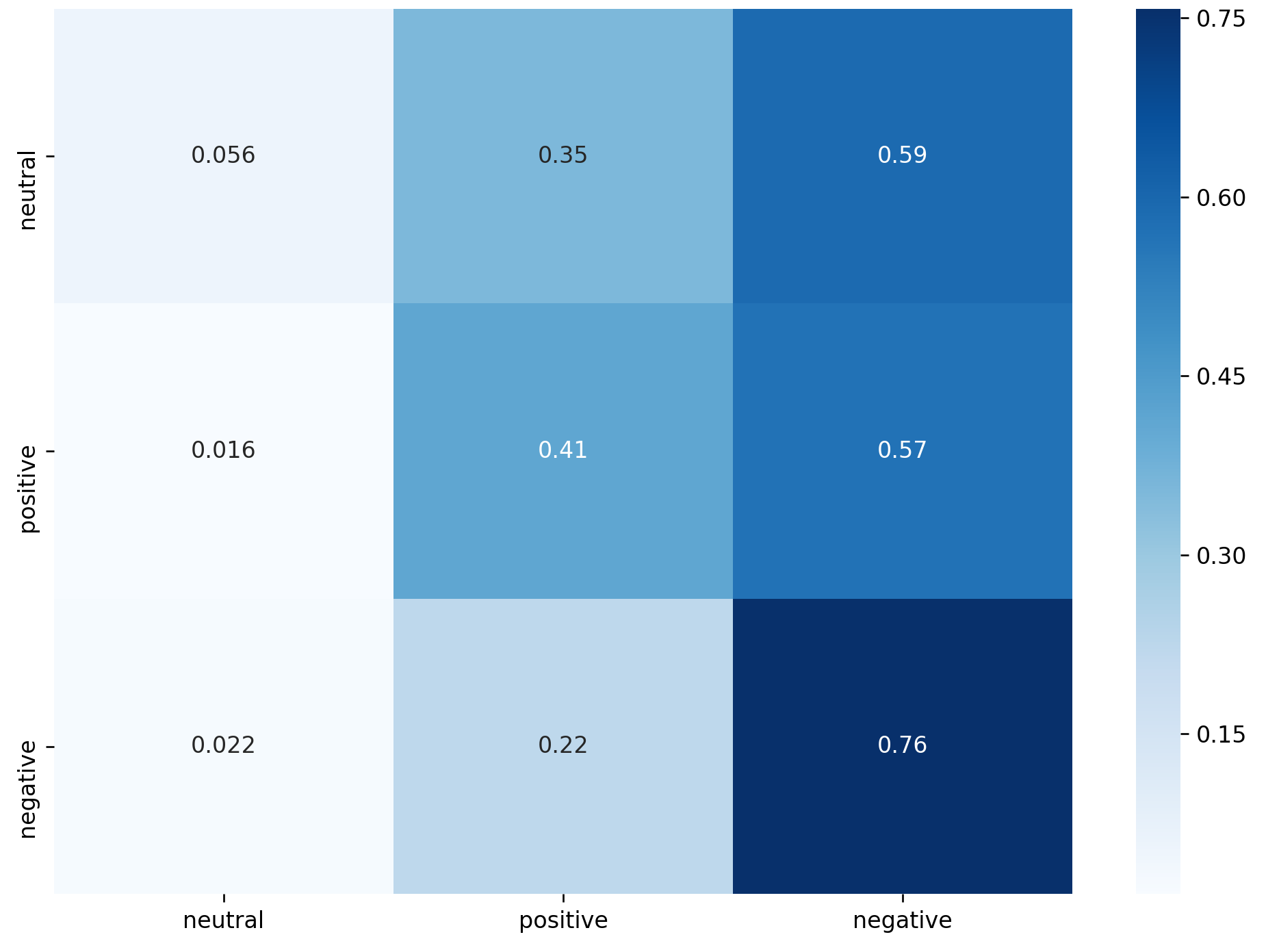}
    \hfill
    \includegraphics[width=0.48\linewidth]{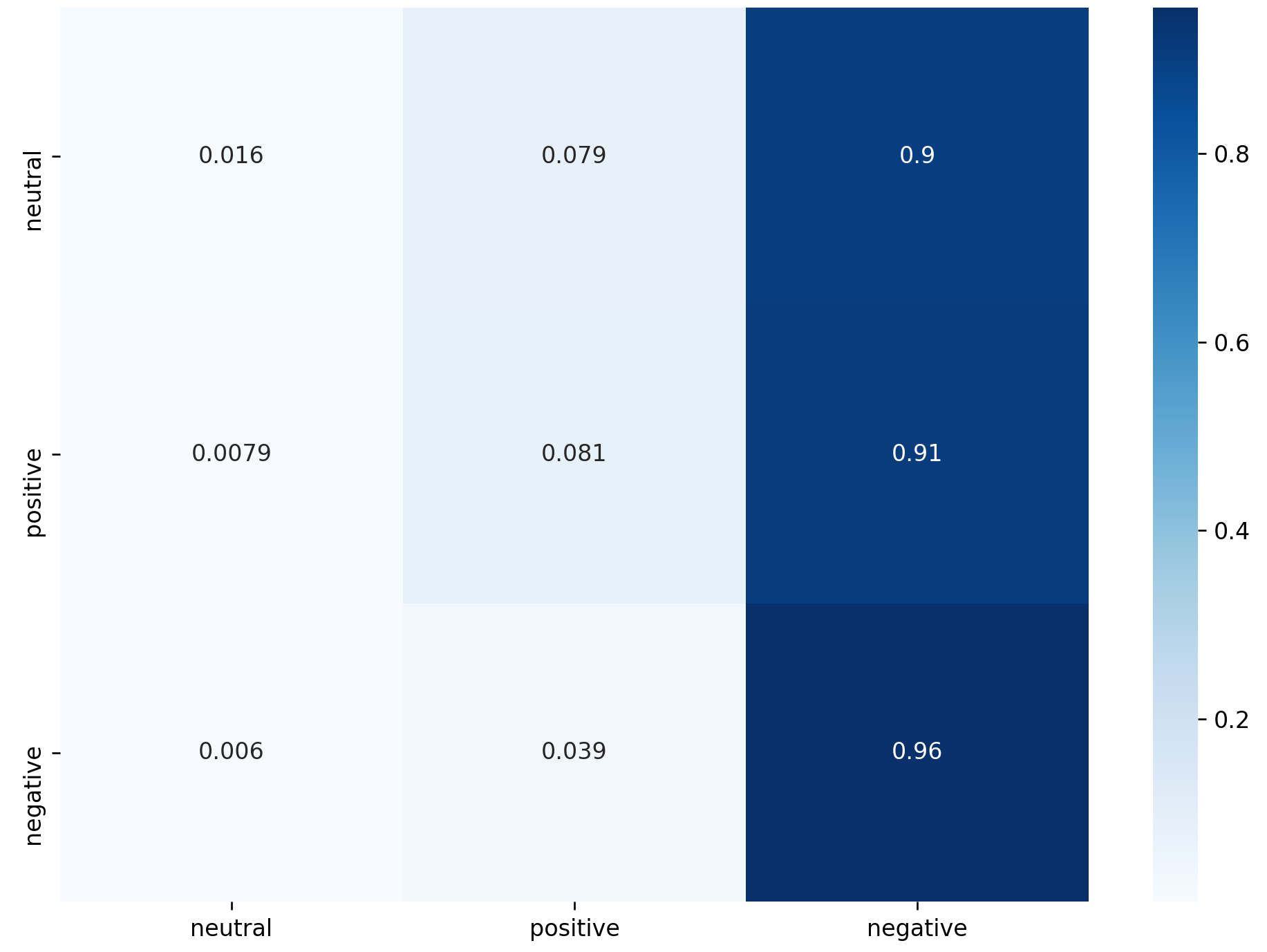}
    \caption{Confusion matrices derived from the three-class cross-corpus experiments with data augmentation for balancing and tests on EMOVO carried out with the end-to-end model with speaker token. The matrices refer to the experiments performed using undersampling (left) and data augmentation (right) as balancing techniques. \label{fig:confusion-matrix-three-classes}}
    \Description{.}
\end{figure}

The effect of this is that the resulting set has thousands more examples for the negative label compared to the other two. Therefore, adding samples to balance the dataset, the models during training observe many generated examples for the positive and neutral classes, while for the negative label the observed signals are all original. As a result, in inference, the models tend to generalize well for the negative class and worse for the other two, with the consequence that most of the speech signals are predicted with the majority class, as shown in Figure~\ref{fig:confusion-matrix-three-classes}.

Finally, we note that Table~\ref{tab:three-classes} also shows the UAR value taken from~\cite{Parry2019Analysis}, one of the few works that proposed a CNN architecture for cross-corpus SER. 
We note that although this value is similar to those obtained from our models, the data used in~\cite{Parry2019Analysis} are partially different from ours. In particular, IEMOCAP, EmoDB, RAVDESS and SAVEE are used in~\cite{Parry2019Analysis} as in our work, but instead of DEMOS the authors in~\cite{Parry2019Analysis} considered the EPST~\cite{Liberman2002EST} dataset. 

\subsection{Additional Discussion of the Results and Future Perspectives}
Among all the experiments carried out, we consider those related to the four-classes task, as reported in the paper, as the most relevant ones.
In relation to this, the performance analysis obtained with each model highlighted the already known difficulty in training cross-corpus models in SER. In fact, the accuracy of the best models are not yet sufficiently high to allow the implementation of a system that can be of concrete help in daily life.
Despite this, the proposed solutions can provide inspiration and indicate possible directions for further investigations.
First of all, it was observed that considering the four-class task, the low accuracy is partly caused by the fact that the predictions of the classes relating to happiness and neutrality are extremely low, compared to high precision for the other two associated with anger and sadness. This could be reduced by balancing the training dataset so to have more samples for the first two labels and fewer examples for the other two. 
In addition, the introduction of the self-attention module within the architectures offers various application opportunities. In fact, this mechanism allows learning the relevance of a set of vectors when one in particular is interpreted. In this work, this approach has been used to detect the dependence of the speaker embedding with respect to features extracted from a spectrogram in a module of attention that receives only two vectors as input. 
However, it is evident how using as inputs other representations extracted from the audio track such as the encoding of the corpus to which it belongs to or the genus of the subject who is speaking, it would be possible to add information at a low computational cost.

\bibliographystyle{ACM-Reference-Format}
\bibliography{sample-base}

\end{document}